\documentclass[aps,prb,preprint,longbibliography,showpacs,groupedaddress]{revtex4}
\usepackage{graphicx}
\usepackage{subfigure}
\usepackage{multirow}
\usepackage{dcolumn}
\usepackage{bm}
\usepackage{color}
\newcolumntype{d}[1]{D{.}{.}{1.2}}
\begin{document}

\title{Prediction of huge magnetic anisotropies of transition-metal dimer$-$benzene complexes}

\author{Ruijuan Xiao, Daniel Fritsch,\footnote{present address: School of Physics, Trinity College Dublin, Dublin 2, Ireland.} Michael D. Kuz'min, Klaus Koepernik,
and Manuel Richter} 

\affiliation{IFW Dresden e.V., PO Box 270116, D-01171 Dresden, Germany} 

\author{Knut Vietze and Gotthard Seifert}

\affiliation{Physikalische Chemie, Technische Universit\"at Dresden,
D-01062 Dresden, Germany}

\date{\today}

\begin{abstract}

Based on numerically accurate density functional theory (DFT) calculations, we
systematically investigate the ground-state structure and the spin and orbital
magnetism including the magnetic anisotropy
energy (MAE) of $3d$- and $4d$-transition-metal dimer benzene complexes (${\rm TM_2Bz}$,
TM = Fe, Co, Ni, Ru, Rh, Pd; Bz = ${\rm C_6H_6}$).
These systems are chosen to model TM-dimer adsorption on graphene or on graphite.
We find that Fe$_2$, Co$_2$, Ni$_2$,
and Ru$_2$ prefer the upright adsorption mode above the center of the benzene molecule,
while ${\rm Rh_2}$ and ${\rm Pd_2}$ are adsorbed
parallel to the benzene plane. The ground state of ${\rm Co_2Bz}$
(with a dimer adsorption energy of about 1 eV)
is well separated from other possible structures and spin states.
In conjunction with similar results obtained by {\em ab initio} quantum chemical
calculations, this implies that
a stable ${\rm Co_2Bz}$ complex with 
$C_{6v}$ symmetry is likely to exist.
Chemical bonding to the carbon ring does not
destroy the magnetic state and the characteristic level scheme of the cobalt dimer.
Calculations including spin-orbit coupling show that the huge MAE of the free Co
dimer is preserved
in the ${\rm Co_2Bz}$ structure. The MAE predicted for this structure
is much larger than the MAE of other magnetic molecules known hitherto,
making it an
interesting candidate for high-density magnetic recording. Among all the other
investigated complexes, only ${\rm Ru_2Bz}$ shows a potential for strong-MAE
applications, but it is not as stable as ${\rm Co_2Bz}$.
The electronic structure of the complexes is analyzed and the magnitude of
their MAE is explained by perturbation theory.

\end{abstract}


\maketitle

\section{\label{sec:introduction}Introduction}

Motivated by the ongoing quest for yet higher-density magnetic data
storage in the context of the rapid advance of information technology,
there is a continued search for nanoscopic magnetic structures with a 
large magnetic anisotropy energy (MAE) density by different experimental
\cite{gambardella03,accorsi06,milios07,mannini09,gambardella09,stillrich09}
and theoretical \cite{ravindran01,strandberg07,mokrousov07,smogunov08,
fritsch08,grunner08,conte08,richter09,zhang09} methods.
The MAE is the energy needed to turn the saturated magnetization
of a system from one direction, usually the ground-state orientation,
to another high-symmetry direction.
Talking for simplicity about a nanoscopic
system, we mean a canonical statistical ensemble of such systems,
i.e. a macroscopically large number of identical non-interacting 
systems at equilibrium with a thermal bath.

Thus, the MAE describes the stability of the magnetization
direction against differently oriented external magnetic fields. In the
simplest case of uniaxial anisotropy with an easy axis, the MAE
is the energy barrier between two opposite equivalent
directions of magnetization with the lowest energy.
A well-known example for this situation is bulk {\em hcp} Co.
Using the bi-stability of
magnetic structures, a bit of information can be stored: for
example, one stable direction of magnetization on a hard disc area may encode
``0'', the other stable direction ``1''.   
It is generally accepted that long-term data storage requires that the
total MAE of each magnetic particle should exceed
40 $kT$,\cite{charap97}
where $k$ is the Boltzmann constant and $T$ is the temperature.

Spin-orbit interaction in a magnetic state is the primary source of 
MAE.\cite{brooks40}
The size of the spin-orbit coupling parameter of a given shell
is a merely atomic property that depends on the atomic number.
Hence, the spin-orbit splitting of atomic, molecular, or band states
is fully determined by the character of these states in terms of 
atomic orbitals.
On the other hand, the MAE as an energy difference depends sensitively
on the particular electronic structure and, thus, on the geometry of the
system.
For instance, bulk {\em hcp} Co shows a moderate MAE of 0.06 meV per atom.
Larger MAE can be obtained in surface-supported structures.
It can be considerably influenced by tuning the coordination and the hybridization
through the choice of substrate and size of the deposited clusters.
For instance, the deposition
of single Co atoms on a Pt(111) surface yields a record MAE of 9 meV per Co 
atom.\cite{gambardella03}

Further reduction of the dimensions leads into the realm of nano-particles
and magnetic molecules. To give an example,
Fe$_4$ organometallic clusters with a propeller-like structure exhibit 
magnetic anisotropy barriers which can be tuned by altering the ligands
and reach up to 1.5 meV per cluster.\cite{accorsi06}
A large anisotropy barrier of 7 meV per cluster,
generated by a deliberate structural distortion of the magnetic
core with the help of bulky organic ligands,
was recently reported for a complex within the Mn$_6$ family of
single-molecule magnets.\cite{milios07}

Isolated magnetic dimers
are the smallest chemical objects that possess a magnetic anisotropy as their
energy depends on the relative orientation between dimer axis and magnetic
moment. Huge MAE values of up to 100 meV per atom were predicted for several
transition-metal dimers (Ti$_2$, Fe$_2$, Co$_2$, Ni$_2$, Zr$_2$, Tc$_2$,
Rh$_2$, Ir$_2$, and Pt$_2$).\cite{strandberg07,seivane07,strandberg08,fritsch08}
However, it is impossible to utilize the huge MAE of dimers
technologically unless they are bound to some medium. Our recent studies 
demonstrate that carbon-based
substrates are suitable for this purpose.\cite{richter09}
Both benzene (Bz) and graphene are ideal support materials that do not spoil
magnetism and the huge MAE of the Co dimer.
The Co$_2$Bz complex has a ground-state structure with $C_{6v}$ symmetry,
in which the dimer is bound perpendicularly to the carbon plane.
This hexagonal environment preserves the two-fold degenerate singly occupied
highest molecular orbital (HOMO) of the free Co dimer, which is responsible 
for the large dimer MAE.\cite{strandberg07}
As a result, Co$_2$Bz 
was predicted to show a magnetic anisotropy of the order of 100 meV per 
molecule.\cite{richter09}
This finding may open a way to enhance the presently available
area density of magnetic recording by 3 orders of magnitude.

The present work has two aims. First, additional detailed results on
Co$_2$Bz will be presented that support the conclusions drawn in 
Ref. \onlinecite{richter09}. Second, related results for a whole series
of dimer-benzene complexes including the 3$d$ and 4$d$ dimers
Fe$_2$, (Co$_2$,) Ni$_2$, Ru$_2$, Rh$_2$, and Pd$_2$ will be shown.
Among these, only Co$_2$Bz and Ru$_2$Bz turn out to be
interesting candidates for a potential strong-MAE application.

Benzene-reacted metal dimers have been explored since the 1980s.
Trevor \textit{et al.} \cite{trevor85} studied the
reaction products of benzene with gas-phase platinum clusters of different
size.
The existence of Fe$_2$Bz was verified with infrared spectroscopy
by Ball {\em et al.} \cite{ball86}
after the reaction of iron atoms with cyclic hydrocarbons in an
argon matrix.
More recently,
Bowen's group performed a series of mass spectrometry and photoelectron 
spectroscopic studies on iron-benzene,\cite{zheng08}
iron-coronene,\cite{li08} cobalt-benzene,\cite{gerhards02} cobalt-pyrene,\cite{kandalam08}
cobalt-coronene,\cite{kandalam07} and nickel-benzene~\cite{zheng05}
cluster anions. The electron 
affinities and the vertical electron detachment energies were extracted
from experimental spectra for the above complexes.
By comparing measured spectra with results of density functional theory (DFT)
calculations,\cite{li08,kandalam07}
they proposed structure models for a part of these clusters.
For example, a half-sandwich ground-state structure ($C_{6v}$ symmetry) was
postulated for ${\rm Fe_2Bz^-}$.~\cite{zheng08}
Their investigations suggest that carbon rings could be a suitable 
template to deposit small transition-metal clusters.
The reaction of Rh$_n^+$
cations with benzene was studied by Berg \textit{et al.}\cite{berg96}
These authors found that for $n=2$ small amounts of Rh$_2$Bz$^+$ were formed,
while the main products were RhBz$^+$ + Rh and Rh$_2$C$_6$H$_4$ + H$_2$.
L\"uttgens \textit{et al.}~\cite{luttgens01}
measured the photoelectron detachment spectra of M$_2$Bz$^-$ (M = Pt,
Pd, Pb) and resolved the electron affinities and ground-state vibration energies
of these complexes. 
By analyzing the vibration frequencies, they postulated a 
perpendicular arrangement of Pd and Pt dimers on Bz ($C_{6v}$ symmetry)
and a parallel coordination between ${\rm Pb_2}$ and
Bz ($C_{2v}$ symmetry).

On the theoretical side, DFT calculations were performed
on several transition-metal benzene systems.
Considering reaction products of iron atoms and benzene
in low-temperature matrices, Parker recently calculated the energy of a number of isomers and
simulated the related infrared spectra.\cite{parker10}
By comparison of 
calculated and measured spectra he concluded that
Fe$_2$Bz is formed at high iron concentrations. If the Fe dimer is assumed perpendicular to the benzene
plane above the center of the carbon ring, the calculated infrared spectrum shows an excellent 
agreement with the experimental data, though the calculation finds a different ground-state isomer.

The ground-state structures of Fe$_2$-coronene,~\cite{li08} Co$_2$-pyrene,~\cite{kandalam08}
and Co$_2$-coronene~\cite{kandalam07} were also studied.
For Fe$_2$-coronene, a ground state with a total spin $S$ = 3 was found with three quasi-degenerate
isomers, where the Fe dimer is oriented either parallel or perpendicular to the
coronene plane.~\cite{li08}
For both Co$_2$-pyrene and Co$_2$-coronene, the ground state was found to be $S$ = 2
with perpendicular orientation of the Co dimer at a position above a peripheral
C-C bridge.~\cite{kandalam07,kandalam08}
Earlier work in this field is due to
Senapati \textit{et al.}~\cite{senapati03} who studied neutral
and cationic ${\rm Fe_2}$-coronene complexes but discussed only parallel adsorption modes.
Also, Rao and Jena studied the geometries and magnetic moments of
neutral and ionic Ni$_n$Bz$_m$ complexes.\cite{rao02_a,rao02_b}
They predicted a parallel adsorption mode
between the Ni dimer and benzene, but it is unclear whether a perpendicular geometry 
was considered or not.
The interaction of benzene with Rh$^+$ and with
${\rm Rh_2^+}$ was investigated by Majumdar \textit{et al.}\cite{majumdar97}
For the physisorbed 
dimer cation, Rh$_2$Bz$^+$, the minimum energy geometry 
has $C_{2v}$ symmetry with
the two rhodium atoms lying horizontally above the benzene at the C-C bridge sites.

Further research activities were devoted to
transition-metal dimers interacting with graphene or with fullerenes.
Interaction of silver and gold adatoms and dimers with graphite or graphene was
studied,\cite{wang03,strange08} and a perpendicular orientation of gold
dimers on graphene was predicted.\cite{strange08}
Also, both structure and spin magnetic
properties of 3$d$ transition-metal adatoms and dimers on graphite were
investigated by Duffy \textit{et al.},\cite{duffy98}
but a possible perpendicular
arrangement of the dimers was not considered. 
DFT calculations for palladium clusters supported on
graphene~\cite{cabria10} and on ${\rm C_{60}}$~\cite{loboda06}
find that the two atoms of an adsorbed Pd dimer are located on bridge 
sites, i.e., on top of C-C bonds. Recently, two detailed
theoretical studies of Fe, Co, and Ni adatoms and dimers
adsorbed on graphene were published:
Johll \textit{et al.}~\cite{johll09}  found that
the most stable structure for all considered
dimers, ${\rm Fe_2}$, ${\rm Co_2}$, and ${\rm Ni_2}$, has a
dimer axis oriented perpendicularly
to the graphene plane and placed at the hole site.
An enhancement of the magnetic
moment for the atom farther from the graphene was predicted,\cite{johll09}
compared with the free dimer.
Cao \textit{et al.}~\cite{cao10} found the same ground-state geometry,
if the density-gradient corrected
functional according to Perdew, Burke, and Ernzerhof~\cite{perdew96}
was used, but they note that partly different results were obtained by
using the local spin-density approximation (LSDA).

DFT calculations do not only allow to predict the ground-state geometry
and spin of a magnetic complex,
but may also provide a basic understanding of the electronic structure and of the
orbital magnetic properties like orbital magnetic moment and MAE.
We are however not aware of any published calculations 
of the magnetic anisotropy of transition-metal dimers
on carbon-based systems except our recent letter, Ref. \onlinecite{richter09}.
In the following, we will demonstrate by DFT calculations that carbon hexagons
are suitable hosts, where adsorbed transition-metal dimers may preserve their
exceptional magnetic anisotropy. 

The investigated TM$_2$Bz complexes are meant to serve as model structures
for the adsorption of transition-metal dimers on the surface of graphite,
on graphene, or on other carbon structures including molecular systems.
For this reason, we only consider so-called physisorption (adsorption without
expelling other atoms, e.g., hydrogen) as opposed to chemisorption that
includes the possibility of de-hydrogenation.\cite{majumdar97}
Most probably, the presented predictions can
only be verified under ultra-high vacuum conditions. Any interaction of
the transition-metal atoms with, e.g., oxygen may deteriorate
the specific structure and the related magnetic state we are focusing on.

The paper is organized as follows. In Sec.~\ref{sec:method}, the
calculation method and computational details are explained. Sec.~\ref{sec:results} 
compiles all results and related discussion:
structure optimization and stability of the
ground states, analysis in terms of the bonding mechanism, the
spin and orbital moments, and the strength of
the MAE. The origin of the huge MAE in some of these molecules is
also explained. Finally, the paper is summarized in Sec. IV.
The appendix contains a description of auxiliary calculations.

\section{\label{sec:method}Method and computational details} 
The DFT calculations were performed with an highly accurate all-electron
full-potential local-orbital scheme (FPLO),\cite{koepernik99}
release 8.00-31.\cite{fplo08}
The code is based on a linear combination of
overlapping nonorthogonal orbitals with a compact support.
The molecular mode of FPLO with free boundary conditions was used.
The presented data were obtained using the
generalized gradient approximation (GGA) 
with a parameterized exchange-correlation
functional according to Perdew, Burke, and Ernzerhof.\cite{perdew96} 
All results were checked against additional calculations using the
LSDA in the parameterization by Perdew and Wang.\cite{perdew92}
The dimer adsorption energy 
calculated by GGA is in all cases about 1 eV smaller than the related
LSDA energy, but both approaches find the same ground-state
structure type.
Also, the ground-state spin obtained with GGA or LSDA is the same
for all systems except Rh$_2$Bz, where LSDA yields a non-magnetic
ground state and GGA yields an $S = 1$ ground state.
In both cases, however, the energies of the
$S =0$ and $S = 1$ states are very close.

The molecular levels were occupied according to a Fermi-Dirac distribution
in order to ensure the convergence of the Kohn-Sham equations.
The presented results were obtained with a broadening
temperature of $T = 100$ K.
The basis set
comprised 3$d$-transition-metal (3$s$, 3$p$, 3$d$, 4$s$, 4$p$, 4$d$, 5$s$),
4$d$-transition-metal (4$s$, 4$p$, 4$d$, 5$s$, 5$p$, 5$d$, 6$s$),
carbon (1$s$, 2$s$, 2$p$, 3$s$, 3$p$, 3$d$), and
hydrogen (1$s$, 2$s$, 2$p$) states. Lower-lying states of the
transition-metal atoms were treated as core states.  

\begin{figure}
\includegraphics[width=0.7\textwidth]{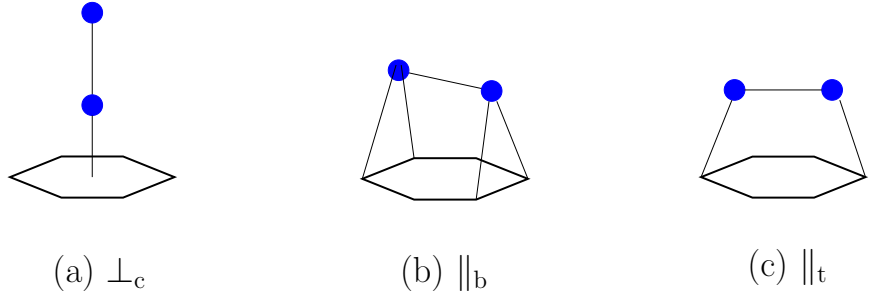}
\caption{\label{fig:str}(Color online) 
Three possible high-symmetry structures for $\rm TM_2Bz$ complexes.
Hexagons and blue bullets indicate benzene rings and and transition-metal
atoms, respectively.
(a) $\perp_{\rm c}$ $-$ the TM dimer is situated on the $C_{6v}$ symmetry axis
perpendicularly to the benzene plane; 
(b) $\parallel_{\rm b}$ $-$ the
TM dimer is parallel to the benzene plane with TM atoms near the middle
of opposite C-C bonds ($C_{2v}$ symmetry);
(c) $\parallel_{\rm t}$ $-$ the TM dimer is parallel to
the benzene plane with TM atoms near the top of opposite C atoms ($C_{2v}$ symmetry).
}
\end{figure}

Geometry optimization was carried out with a scalar relativistic
scheme.
To find out the lowest-energy geometry and spin magnetic state
of each TM$_2$Bz complex, three
possible high-symmetry structures (Fig.~\ref{fig:str}) 
were optimized for $S$ = 0, 1, 2, and 3 (total spin moment, $\mu_{S}$ =
0, 2, 4, and 6$\mu_{B}$), and for different initial spin arrangements
(ferro- and ferrimagnetic). The point group symmetry $C_{6v}$ was applied
for the configuration shown in Fig.~\ref{fig:str}(a), 
while for the structures depicted in
Figs.~\ref{fig:str}(b) and \ref{fig:str}(c),
$C_{2v}$ was used for the structure optimization.

Previous theoretical investigations of metal-benzene systems
have shown that the structural changes of the benzene plane due to the
metal-benzene 
interaction are negligible.\cite{mokrousov07,pandey00,pandey01,philpott07}
Thus, in nearly all our calculations the positions of the C and H
atoms were fixed with C-C bond length 1.40 \AA ~and
C-H bond length 1.09 \AA{}.
Exceptions from this strategy are reported below.

The Co$_2$Bz complex is of particular interest.\cite{richter09}
To make sure that the correct ground state was found 
for this system, we 
utilized the pseudopotential code ESPRESSO-4.0.1~\cite{pwscf} and
cross-checked the
above calculations by full optimization of all the atomic positions
starting from 14 kinds of initial structures. The computational details and 
a brief description of the results are given in the appendix.

A quantity used to judge the stability of the considered
structures is the adsorption energy ($E_{ad}$) for a dimer entity attached to
a benzene molecule, which is defined as
\begin{equation}
E_{ad}= E_{tot}{\rm (Bz)} + E_{tot}{\rm (TM_2)} -E_{tot}{\rm (TM_2Bz)} \; .
\end{equation}
Here $E_{tot}$ refers to the respective total energy of the species
indicated in the parentheses.
Negative values of $E_{ad}$ mean that the TM$_2$Bz complex is unstable.

To evaluate the orbital magnetic moment and the MAE,
spin-orbit coupling has to be included in the calculation.
However, standard (quasi-)local DFT approximations like LSDA
or GGA do not include orbital-dependent exchange effects.\cite{eschrig05}
Thus, the orbital moments and the MAE
are usually underestimated by these approaches.
The orbital polarization (OP) correction~\cite{eriksson90} is a frequently 
applied method to cure this
problem. As a matter of experience, the MAE evaluated with
standard LSDA or GGA approximation gives a lower estimate to the expected MAE,
while the value obtained by including the OP correction provides an upper
estimate. Experimental values of the MAE are most probably
located between these lower and upper estimates.
This has been demonstrated, e.g.,
in Refs. \onlinecite{trygg95} and \onlinecite{ravindran01}
and also for the special case of Co atoms in different chemical
and structural surroundings in Ref. \onlinecite{richter09}, Fig. 3.

The MAE and the orbital magnetic moment were calculated by means of
self-consistent fully relativistic calculations using the bond lengths obtained
in the scalar relativistic calculations.
The MAE was defined as
\begin{equation}
\label{eq:2}
{\rm MAE} = E_{tot} [\parallel ] -  E_{tot} [\perp ] \; ,
\end{equation}
where $E_{tot} [\parallel ]$ and $E_{tot} [\perp ]$ denote total
energies of states with magnetization direction parallel and perpendicular
to the Bz plane, respectively.
The choice of the direction parallel to the plane is arbitrary, since the
in-plane anisotropy is negligible on the scale of the considered energies.
Results obtained with and without OP corrections are reported.
In the former case, the spin-dependent OP correction~\cite{nordstroem92}
was applied to the 3$d$-orbitals of Fe, Co, Ni and
to the 4$d$-orbitals of Ru, respectively.

In order to cross-check one of the most important details of the GGA
calculations, the bonding behavior of the Co dimer with Bz,
we also performed {\em ab initio} quantum chemical calculations
at the level of second order M\o{}ller-Plesset perturbation theory (MP2).
The MP2 results were obtained from the MP2 implementation of Gaussian03.~\cite{Gau}
For Co$_2$Bz, the Co atoms were described with a scalar-relativistic effective
core potential (ECP) replacing 10 core electrons (MDF10),~\cite{Dolg} with the
corresponding ($8s7p6d$ $1f$)/[$6s5p3d$ $1f$] GTO basis set
of triple-zeta quality.
Accordingly, for benzene the Dunning correlation-consistent basis sets of
double- and triple-zeta quality (cc-pVDZ and cc-pVTZ, respectively)~\cite{Dun} were used.
Since all of these basis sets are rather large, all energies have been corrected for
the basis set superposition error (BSSE) with respect to the dissociation of the Co
dimer from the benzene ring, employing the counterpoise scheme
proposed by Boys and Bernardi~\cite{Boys,Sim} as implemented in Gaussian03.

\section{\label{sec:results}Results and discussion} 
\subsection{\label{sec:structural}Structure and spin state}
Dimer adsorption energies for each considered total spin and symmetry
are shown in Fig.~\ref{fig:eb}.
The optimized structure parameters, dimer adsorption energies,
and related spin magnetic moments for 
the six ground-state structures are listed
and compared with literature data
in Table~\ref{tab:eb}.

\subsubsection{3$d$ transition-metal complexes}
\begin{figure} 
\includegraphics[width=1.0\textwidth]{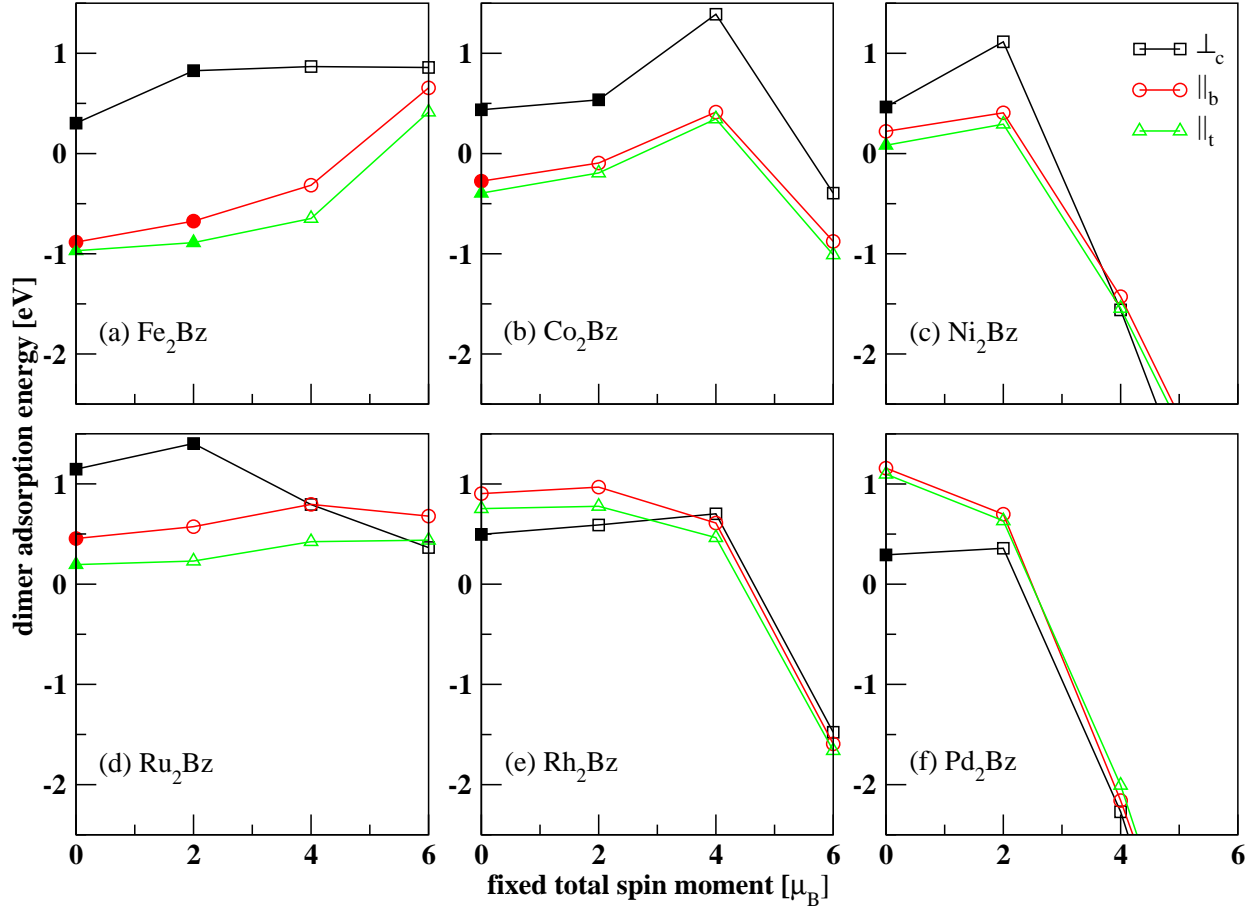} 
\caption{\label{fig:eb}(Color online) Scalar-relativistic 
dimer adsorption energies $E_{ad}$ calculated for optimized
structures of (a) $\rm Fe_2Bz$, (b) $\rm Co_2Bz$, (c) $\rm Ni_2Bz$, (d) $\rm
Ru_2Bz$, (e) $\rm Rh_2Bz$, and (f) $\rm Pd_2Bz$ complexes. For all systems,
three initial
structures illustrated in Fig.~\ref{fig:str} were optimized
with fixed C and H coordinates
for the following
values of the fixed total spin moment, $\mu_S$=0, 2, 4, and 6$\mu_B$. Both
parallel and anti-parallel relative spin orientations were considered.
Open (filled) symbols indicate that the moments of the two
transition-metal atoms in the spin state with the highest adsorption energy
are parallel (anti-parallel).
}
\end{figure}

We find that the adsorption mode
with the dimer axis perpendicular to the benzene plane results in the
most stable structure for all investigated $3d$ systems, Fe$_2$Bz, Co$_2$Bz, and Ni$_2$Bz.
This ground-state geometry is consistent with other GGA results 
obtained for dimers on graphene.~\cite{johll09,cao10}
The only discrepancy occurs for the ground-state
spin magnetic moment of Fe$_2$Bz.
Here, we found a reduction of the free dimer spin ($S$ = 3)
to $S$ = 2, whereas Johll \textit{et al.} and Cao \textit{et al.} had reported $S$ = 3
for the marginally different situation of
Fe$_2$ on graphene.\cite{johll09,cao10}
As the related energy difference in our calculation was very small
(9 meV), we repeated the calculations with full optimization of the
C-C and C-H distances ($C_{6v}$ symmetry). The full optimization
inverted the order of the two considered states, the state
with $S$ = 3 now being 16 meV lower than the competing state
with $S$ = 2.
Moreover, a ferrimagnetic state with $S$ = 1 is found only about 40 meV
higher in energy, Fig.~\ref{fig:eb}. 
Such small energy differences
cannot guarantee the stability of the Fe$_2$Bz magnetic ground state
and should give rise to strong spin fluctuations.

Turning our attention to the ground-state geometry of Fe$_2$Bz, we note that
Parker~\cite{parker10} provided evidence of the same proposed geometry 
by obtaining an excellent agreement between calculated
and experimental infrared spectra. Note, that in this geometry Fe$_2$Bz complexes with
$S$ = 1, 2, and 3 give almost identical simulated spectra.\cite{parker10}
Thus, the comparison cannot be used to distinguish the magnetic state.
Our results add additional weight to Parker's arguments, who proposed
a $C_{6v}$ geometry for Fe$_2$Bz.
One should note, however, that other structure types compete with the
$C_{6v}$ geometry, see Fig.~\ref{fig:eb}. Indeed, if the C and H coordinates
are optimized as well, the ground state turns to $\parallel_{\rm b}$, $S$ = 3,
almost degenerate with the states $\perp_{\rm c}$, $S$ = 2 and $S$ = 3.
In line with this finding, Cao {\em et al.} reported a ground state with the 
Fe dimer above a graphene hollow site, but not perpendicular to the 
plane.\cite{cao10} In the following discussion, we will disregard 
Fe$_2$Bz structures different from $C_{6v}$ due to the mentioned 
experimental evidence of this structure type.

For Co$_2$Bz and Ni$_2$Bz, we obtained both the magnetic moment 
and the ground-state structure in agreement with the results 
by Johll \textit{et al.}~\cite{johll09} and by Cao \textit{et al.}~\cite{cao10}
The GGA dimer adsorption energy,
$E_{ad}$ = 1.39 eV for Co$_2$ on Bz from our calculation,
and the related energies 0.92 eV from Johll \textit{et al.}
and 1.13 eV from Cao \textit{et al.} for Co$_2$ on graphene
indicate a reasonable stability of this structure.
Noteworthy, any other spin state considered
in our calculations, including a ferrimagnetic solution with
a total spin $S$ = 1, has a much higher energy, at least 0.80 eV above the 
ground state. The ground state of Ni$_2$Bz is also sufficiently
separated from other states with a different spin and/or geometry, see
Fig.~\ref{fig:eb}. 
We also checked a further geometry with Ni atoms close
to next nearest C-bridges (see Fig. A1 (iv)). Such a geometry was
found by Rao and Jena~\cite{rao02_b} to be lowest in energy.
We could not confirm this finding and obtained, even with full
relaxation, a 0.45 eV higher energy for this geometry ($S$ = 1) than for the 
$\perp_{\rm c}$ geometry with $S$ = 1.

To investigate the stability of the proposed perpendicular adsorption mode
for the Co system in more detail, three further structures were considered,
including two dissociated cases: (i) attachment of one Co atom on
each side of the carbon ring, (ii) dissociation of one of the Co atoms
resulting in one free Co atom and CoBz, and (iii) dissociation of both Co
atoms resulting in two free Co atoms and one free benzene molecule.
In the related scalar relativistic atom calculations,
non-integer occupation of the open shells was admitted.
All the three structures have higher energies, by 2.47 eV, 3.49 eV, and 5.05 eV,
respectively, than the perpendicular arrangement. 
From the energy difference between state (ii) and state (iii), 
a value of 1.56 eV is found for the adsorption energy of a single Co adatom on benzene.
We also can find a binding energy of 3.66 eV for the Co dimer from the
third dissociated state. 
In comparison with the DFT data, the experimental binding or
adsorption energies are considerably
smaller: 0.34 eV~\cite{kurikawa99} (adsorption energy of a Co adatom to benzene)
and 1.72 eV~\cite{lombardi02} (binding energy of a Co dimer).
This is in line with the known tendency of 
DFT calculations to overestimate the binding energies in many cases.

\begin{figure}
\includegraphics[width=0.84\textwidth]{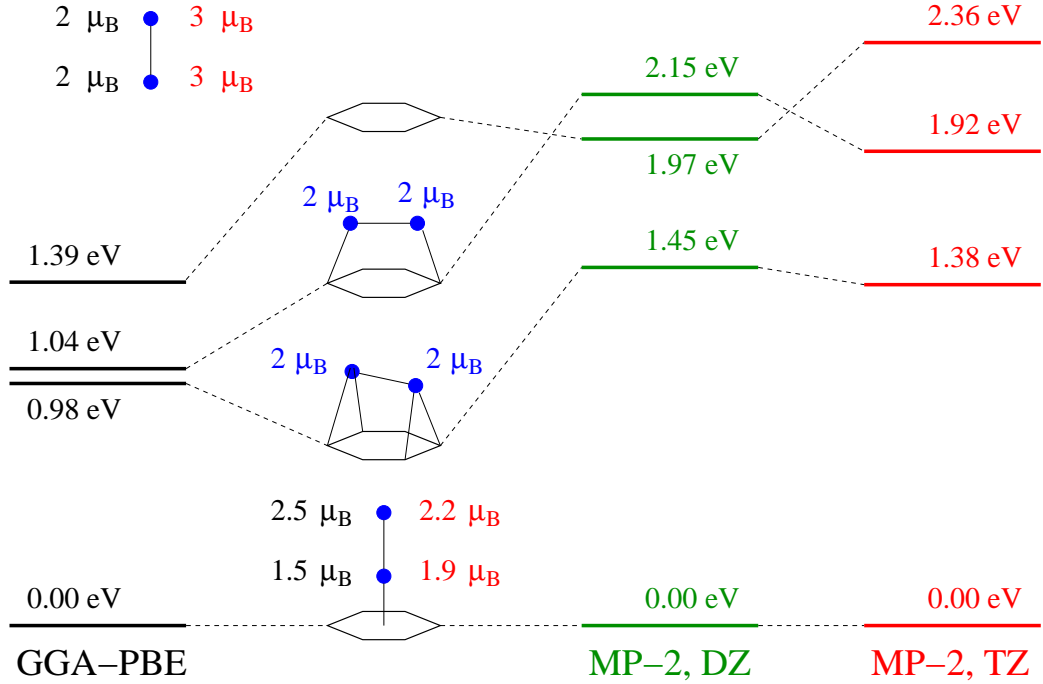}
\caption{\label{fig:gga-mp2}(Color online) Energies and magnetic moments 
of different Co$_2$Bz configurations calculated by DFT (left column) and by MP2
(right columns).
DZ and TZ abbreviate double-zeta and triple-zeta basis sets, respectively.
The energies refer to the ground-state energy.
For the uppermost, dissociated and for the lowermost,
ground-state configuration, the
spin moments in red (on the right-hand side of the Co dimer) refer
to the MP2 calculations and the moments in black
(on the left-hand side of the dimer) refer to the GGA calculations.
For the other configurations, GGA and MP2 yield the
same spin.
Hexagons and blue bullets indicate Bz and Co, respectively.}
\end{figure}

Thus, to confirm the qualitative validity of the energies
and structure sequence, we performed quantum chemical (MP2) calculations.
In comparison with our previously published results,~\cite{richter09}
the MP2 calculations were improved by taking into account
BSSE corrections and by extending the benzene basis.
Fig.~\ref{fig:gga-mp2} compares
the energy sequences of three
possible high-symmetry structures and one
dissociated configuration obtained by GGA with related MP2 results.
All MP2 energies were evaluated by single point calculations using the 
GGA-derived geometries, except for the Co dimer. For the latter the interatomic 
distance was optimized at the MP2 level. It turned out slightly shorter 
(0.1909 nm) than the GGA result (0.1997 nm). 
For all structures, MP2 calculations were carried out for
$S$ = 0, 1, 2 and 3. In all cases a total spin
of 2 was found to be most favorable.
Importantly, the adsorption energy of the Co dimer to benzene was found to be 
yet higher than in the GGA calculation.
It is obvious from Fig.~\ref{fig:gga-mp2} that the 
quantum chemical calculations confirm the main GGA result,
that bonding of a Co dimer with a single molecule of benzene results in the
structure depicted at the bottom of Fig.~\ref{fig:gga-mp2} with a total spin
$S$ = 2.
Also, the sequence of the higher-energy structures
is the same in GGA and in MP2, and the same spin magnetic moments 
are found with the exception of the dissociated state, where MP2
predicts $S$ = 3 and GGA yields $S$ = 2.
When improving the benzene basis from a double- to a triple-zeta level, the
BSSE-corrected adsorption energies rise consistently by about half an eV, reaching
2.36 eV for the most stable structure. This result should
serve as a valid proof that the Co dimer can be bound to the benzene ring.

As a final check that the calculations described above provided
the correct ground-state geometry, we performed a cross-check
for Co$_2$Bz with the pseudopotential code ESPRESSO-4.01.\cite{pwscf}
We carried out a full
optimization of all atomic positions starting from 14 kinds of initial
structures.
As before, GGA and a scalar relativistic mode were used.
Other technical details are described in the appendix.
The results confirm that the bonding of ${\rm Co_2}$ with a single molecule of
benzene very likely results in the perpendicular configuration, which is
separated from other possible arrangements by at least several hundred meV.
Only a tiny distortion of the benzene plane is found in the full
optimization results. This is a weak Jahn-Teller effect that splits the singly
occupied two-fold degenerate HOMO state originating from ${\rm Co_2}$.
This splitting is very small ($<$1 meV), since the original
$C_{\infty}$ symmetry of the Co dimer where the 
HOMO resides is only weakly distorted by the hexagonal ligand.
Thus, it will hardly have any influence on the
magnetic properties of Co$_2$Bz. 
In particular, if spin-orbit
coupling is taken into account, the HOMO will be split by this interaction
rather than by the Jahn-Teller effect, which is almost 2 orders of magnitude
weaker than the spin-orbit coupling in the considered case.
Only if the magnetization is oriented perpendicular to the
dimer axis, the spin-orbit splitting vanishes in lowest order.
In this case, the
Jahn-Teller effect might marginally reduce the total energy and, thus,
the MAE.

We conclude that Co$_2$Bz and Ni$_2$Bz probably exhibit a $C_{6v}$
symmetry, like Fe$_2$Bz. 
As distinct from the Fe system, Co$_2$Bz and Ni$_2$Bz
have a stable magnetic ground state.
Experimental evidence of these proposed structures seems however lacking
at the moment. 
For the anion Co$_2$Bz$^-$, 
the observed photoelectron spectra~\cite{gerhards02}
allowed to exclude
a structure where the two Co atoms are
placed on both sides of the benzene plane.

\subsubsection{4$d$ transition-metal complexes}
We find (Fig. \ref{fig:eb}) that among the investigated 4$d$ dimers
only $\rm Ru_2$ prefers an upright
adsorption mode. It binds to the benzene as strongly
as the cobalt dimer, but its magnetic ground state has a
lower spin, $S$ = 1, and lies only 0.25 eV below a zero
spin state. The ground states
of both Rh$_2$Bz and Pd$_2$Bz are found to be almost degenerate with
respect to spin multiplicity (Rh$_2$Bz) or geometry (Pd$_2$Bz).

We are not aware of any published information about
the geometry of the neutral complexes Ru$_2$Bz and Rh$_2$Bz.
For the cation Rh$_2$Bz$^+$, the structure type $\parallel_{\rm b}$
(Fig. \ref{fig:str}) was obtained as the lowest-energy structure
by DFT calculations using the B3LYP functional.\cite{majumdar97}
This is the same ground-state structure as we find for the neutral Rh$_2$Bz.

L\"uttgens \textit{et al.}~\cite{luttgens01} deduced the vibration energies of
both Pd$_2$Bz and Pd$_2$Bz$^-$ from photoelectron detachment spectra.
They postulated an orientation
of the Pd dimer perpendicular to the benzene ring
because the observed vibration frequency of Pd$_2$Bz is close to that 
of the free Pd$_2$. One should note that our calculated ground-state geometry 
of Pd$_2$Bz contradicts this analysis.
On the other hand, calculations by 
Cabria \textit{et al.}~\cite{cabria10} using LSDA and GGA and by 
Loboda \textit{et al.}~\cite{loboda06} using the B3LYP functional
find a ground-state geometry with the two palladium atoms placed
horizontally above the carbon ring, similar to our results.

We performed a series of additional tests in order to clarify this
discrepancy between experiment and theory.
First, we checked the influence of spin-orbit interaction
and found, that related total energy shifts do not exceed 0.1 eV.
The parallel adsorption mode hence is still more stable than 
the perpendicular one. Second, we checked a possible asymmetric
adsorption above a single bridge site.
Indeed, the related total energy is about
0.5 eV lower than for the adsorption above the hollow site, if every
symmetry constraint is released.
The dimer axis then deviates from the initial perpendicular orientation
and forms an angle of about 40 degrees with the benzene plane, while $S=0$.
Yet, the energy of this structure is still higher than that of
the parallel configuration. Third, we calculated vibration frequencies
of the Pd-Pd bond using the harmonic approximation for the
three adsorption modes defined in Fig.~\ref{fig:str}.
If $S$ = 1, the three structures give rise to almost the same
vibration energies of 21.9 meV ($\perp_{\rm c}$), 22.3 meV ($\parallel_{\rm b}$)
and 22.5 meV ($\parallel_{\rm t}$).
For the asymmetric bridge site configuration, a value of 20.6 meV is obtained.
All of these energies are close to the vibration
energy of the free dimer, 26 meV.\cite{ho93}
This comparison shows that proximity of the vibration spectrum
to that of the free dimer is no proof of the perpendicular geometry.
For the ground state, $S$ = 0,
the parallel structures turn out to be much softer, 12.8 meV ($\parallel_{\rm b}$)
and 13.3 meV ($\parallel_{\rm t}$), than the free dimer and also than the 
perpendicular geometry, 19.7 meV.
On the whole, the calculated vibration spectra do not provide enough evidence 
in favor of any investigated structure.
Finally, we optimized the structures 
of Pd$_2$Bz$^-$ anions. The structure $\parallel_{\rm b}$ with 
$S$ = 1/2 is again more stable than the structure $\perp_{\rm c}$, by 0.45 eV.
The structure with asymmetric bonding above a single bridge site
is 0.13 eV lower in energy than the structure $\perp_{\rm c}$, but it is
still 0.32 eV higher than the structure $\parallel_{\rm b}$.
Summarizing this point, the discrepancy between experimental and
theoretical results on Pd$_2$Bz persists.

\subsection{\label{sec:bonding}Electronic structure and
bonding mechanism} 
Free TM dimers have been discussed in detail 
recently.\cite{strandberg07,strandberg08,fritsch08,blonski09}
Analysis of their electronic structure
reveals that a singly occupied HOMO which is two-fold degenerate
in the absence of spin-orbit coupling is responsible
for the giant magnetic anisotropy predicted in some of these 
dimers.\cite{strandberg07}
For example, the most important feature
in $\rm Co_2$ is a two-fold degenerate singly occupied 3$d$-$\rm \delta_u^*$
state. It is split by spin-orbit interaction, if the magnetic moment is
oriented along the dimer axis but stays degenerate if the moment is oriented
perpendicular to the axis.\cite{strandberg07}
Concerning the bonding between metal atoms
and a benzene molecule, Mokrousov \textit{et al.}~\cite{mokrousov07} reported a
schematic analysis for V-Bz complexes and showed that the HOMO and the lowest
unoccupied molecular orbital (LUMO) of benzene interact with the metal $s$ and $d$
orbitals of the same symmetry.

To better understand the bonding mechanism
between the TM dimer and the benzene molecule, we compare the levels
of Co$_2$Bz and of Fe$_2$Bz
with those of the related free dimers, Fig.~\ref{fig:level}.
The third panel (from left) shows the textbook electronic structure
of benzene and the leftmost panel refers to the free $\rm Co_2$,
as recently discussed in Ref.~\onlinecite{strandberg07}. 
The other panels show the electronic structure evaluated
for the ground-state geometries 
and spin multiplicities of $\rm Co_2Bz$, Fe$_2$Bz, and Fe$_2$
as well as for the low-lying $S$ = 3 state of Fe$_2$Bz.
It turns out that bonding of $\rm Co_2$ on benzene does not lead
to any deterioration of the magnetic properties of $\rm Co_2$: (i) the 
ground-state spin stays $S$ = 2, as in the free dimer; (ii)
the Co 3$d$-$\rm \delta_u^*$ level is still
two-fold degenerate in Co$_2$Bz due to the $C_{6v}$ symmetry; (iii)
this level is still the singly occupied HOMO.
In this way, the key feature responsible for the giant MAE of the
free Co dimer is preserved in the Co$_2$Bz
structure if the benzene molecule binds perpendicularly to Co$_2$ in 
$C_{6v}$ symmetry. 
Yet, there is an important difference between free Co$_2$ and Co$_2$Bz.
In the free dimer, the two Co atoms contribute equal weights to 
the minority spin 3$d$-$\rm \delta_u^*$ state.
At variance, the HOMO of $\rm Co_2Bz$ receives 94\% of its weight
from that Co atom which is farther away from the benzene
plane, see Fig. \ref{fig:orbcomp}.

One can note that the magnetic moment
of free Co$_2$ is also preserved in the parallel arrangements $\parallel_{\rm b}$ and 
$\parallel_{\rm t}$ (Fig.~\ref{fig:eb}). However,
the reduction of the symmetry to $C_{2v}$
causes a split of all $\delta$ and $\pi$ states
(not shown here), resulting
in non-degenerate HOMO and LUMO in these structures.
The minority spin $\pi^*$ state near
to the Fermi level is split by 0.57 eV in the
$\parallel_{\rm b}$ structure and by 0.60 eV
in the $\parallel_{\rm t}$ structure, while the $\delta^*$ state
in the minority spin channel is split yet more strongly,
by 0.68 eV in $\parallel_{\rm b}$ and by 0.85 eV in $\parallel_{\rm t}$.

We find that,
unlike in the case of Co$_2$, the adsorption of Fe$_2$ on benzene
results in a change of both magnetic moment and electron 
configuration, as compared with the free Fe dimer.
The ground-state level scheme
of Fe$_2$ is very similar to that
of Co$_2$, see Fig. \ref{fig:level}, but two more holes are introduced
in the minority spin channel. As a result,
$S$ = 3 (Fe$_2$) instead of $S$ = 2 (Co$_2$) and the
exchange splitting is enhanced, so that the majority spin
$4s\sigma^*$ level is again quasi-degenerate with the HOMO.
The latter is now allocated to the singly occupied $\delta$ orbital
instead of $\delta^*$ in Co$_2$.
If Fe$_2$ binds to benzene, the electron occupying the
majority spin $\pi^*$ level of Fe$_2$ moves into the minority spin
$\delta$ level.
The minority spin $\delta$ level becomes
doubly occupied, while the majority spin $\pi^*$ level turns singly occupied and  
acts as the HOMO in the $S$ = 2 ground state of Fe$_2$Bz.

The electronic structure of Fe$_2$Bz 
is very similar to that of Co$_2$Bz in the same spin state ($S$ = 2).
The covalent splittings of the $d$-states are somewhat larger
in the Fe system than in the Co one,
due to the larger extension of the Fe-$d$ orbitals
compared with the $d$-orbitals of Co.
This yields a somewhat different orbital order.
It will be recalled that in Fe$_2$Bz a state with
$S$ = 3 is close in energy to the ground state; in this state the
exchange splitting is larger and both holes enter the
minority spin channel, like in the free Fe dimer.
We included the corresponding level scheme for comparison
in Fig. \ref{fig:level}.
The essential difference between Fe$_2$ and excited Fe$_2$Bz in the same
spin state is that $\sigma$ and $\delta$ in the minority spin channel
interchange their positions. Thus, Fe$_2$Bz in the state $S$ = 3
has a fully occupied $\delta$ HOMO and, thus, a small magnetic
anisotropy.

The orbital characteristics at the Fermi level determine the main
physical properties of the
structure, while the stability of the complex depends on how the TM
dimer binds to the benzene molecule. Fig.~\ref{fig:orbcomp} 
shows the integrated density of states (IDOS)
for the ground state of the ${\rm Co_2Bz}$ structure
and the orbital composition of each state.

In an axial symmetry,
the five $d$ orbitals are split into three
groups: $d\sigma$($d_{z^2}$), two-fold
degenerate $d\pi$($d_{xz}$, $d_{yz}$), and two-fold degenerate
$d\delta$($d_{xy}$, $d_{x^2-y^2}$).
The six benzene $\pi$ orbitals can also be
classified with respect to the same axis~\cite{yasuike99} and
include one $L\sigma (\pi_1$) orbital, two degenerate HOMO
$L\pi (\pi_2,\; \pi_3)$, two degenerate
LUMO $L\delta (\pi_4^*,\; \pi_5^*)$, and one $L\phi (\pi_6^*)$
orbital, where $L$ means ligand. 
The adsorption between the Co dimer and the benzene molecule is
realized primarily by forming three types of
chemical bonds, $\delta$, $\pi$, and $s\sigma$,
between carbon atoms and Co1 (the Co atom nearest to the benzene ring),
while the the other Co atom, Co2, mainly binds with Co1 and scarcely
contributes to the bonding with the benzene. For example,
the minority spin $\delta$ state at -4.3
eV and the related $\pi^*$-dominated (Bz-$L\delta$) $\delta$-state
at -1.2 eV in Fig.~\ref{fig:orbcomp} stem from the combination of the $d_{xy}$, 
$d_{x^2-y^2}$ orbitals of Co1 and the LUMO of benzene.
The $\pi$ states positioned near -7.8 eV in
the minority spin channel and around -8.0 eV in the majority channel
can be attributed to the
$d_{xz}$, $d_{yz}$ orbitals of Co1 and the HOMO of benzene.
The hybridization between the 4$s$ orbital of Co1 and the
$L\sigma (\pi_1)$ orbital of benzene forms the $s\sigma$ orbital,
located at around -10.2 eV.
Formation of these three kinds of bonds
lowers the energy of the $\rm
Co_2Bz$ complex compared with the dissociated state
and stabilizes the perpendicular adsorption structure.

A simple model for $\rm Co_2Bz$ can be sketched from above analysis of the bond
mechanism: the benzene molecule plays the role of a substrate to fix the
Co dimer, the Co atom next to the benzene plane acts as ``glue'' to
bind the dimer to the benzene, and the other Co atom mainly contributes to the
states near the Fermi level and dominates the 
magnetic properties of the whole structure.  

Level schemes for the ground states of Ni$_2$Bz, Ru$_2$Bz and for the related
free dimers/benzene are shown in Fig. \ref{fig:levelniru}.
The electronic structure of Ni$_2$ is almost unchanged upon adsorption.
The spin remains $S$ = 1
and $\pi^*$ of the minority spin channel is still the singly occupied
HOMO. The situation is different for Ru$_2$, where the total spin
is reduced from $S$ = 2 to $S$ = 1 due to the adsorption.
Together with the change of spin, an essential alteration of the orbital
order close to the Fermi level takes place.
In particular, majority spin $\sigma^*$ and $\pi^*$ levels interchange
their positions. Thus, Ru$_2$Bz has a singly occupied $\delta^*$ HOMO
like Co$_2$Bz.

\subsection{\label{sec:spin}Spin and orbital moments}
Site-resolved spin moments $\mu_S$ and orbital moments $\mu_L$ for the
stable magnetic structures with perpendicular geometry (TM = Fe, Co, Ni, Ru)
are listed in Table~\ref{tab:mag}.
The calculations were carried out within the fully relativistic scheme.
The moments are aligned
perpendicularly ($\perp$) or parallelly ($\parallel$)
to the benzene plane, respectively. 
We checked that the in-plane anisotropy of the magnetic moments
is marginally small and can be safely neglected.
The effect of the
OP correction is also given for comparison. It is clear that the magnetic moment is
mainly distributed between the two transition-metal atoms.
The magnetic moment on the C sites is
so small that it can be neglected ($|\mu_{S(C)}| < 0.05 \mu_B$). 

Inspection of the $\mu_{S}$ data reveals that
in all four systems the TM2 atom shows a much higher spin
moment than the TM1 atom. 
The relatively high coordination number
of TM1, seven, results in a considerable reduction of its spin moment.
On the other hand, TM2 is only singly coordinated and thus behaves
almost like a free atom.
The spin moments of Fe, Co, Ni, and Ru free atoms amount
to 4, 3, 2, and 4 $\mu_{B}$, respectively. 
The calculated spin moments
of the TM2 atoms in $3d$-TM$_2$Bz complexes are only $0.5 \ldots 0.7 \; \mu_B$
smaller than the corresponding atomic values.
In the case of Ru, the TM2 carries only about half of the atomic 
spin moment. Spin magnetism of
$4d$ atoms is in general less stable than that of isoelectronic $3d$ atoms,
since the $4d$ intra-atomic exchange (Stoner) integrals are somewhat smaller
than the related $3d$ integrals.

Another feature is that the
spin moments are nearly the same for both magnetization directions. When the
moment orientation switches from $\perp$ to $\parallel$,
the primary effect is a change of
the orbital moment of the TM atoms. This fact indicates that the magnetic
anisotropy of these systems is closely connected to
the anisotropy of the orbital moments.\cite{bruno89}
It is worth noting that in both the $\rm Co_2Bz$ and the $\rm Ru_2Bz$ systems
the TM2 atoms show very large orbital moments in the $\perp$ orientation and
relatively small values in the $\parallel$ orientation.
This is a sign of a large magnetic anisotropy of these systems.

As expected, the calculated moments in Table~\ref{tab:mag} show that the
OP correction generally increases the orbital moment while 
scarcely affecting the spin
moment. For example, when the OP correction is allowed for, the orbital
moments of Co atoms with magnetic moments parallel to the Bz
plane are about three
times larger than those calculated without the OP correction.
This is caused by the very construction
of the OP correction scheme, where additional (exchange)
energy is gained if the orbital
moment is enhanced.\cite{eriksson90,eschrig05}
In case of perpendicular orientation of the moments with 
respect to the Bz plane, the orbital moments are less
influenced by the OP correction, since spin-orbit coupling
alone already provides almost the maximum orbital moment
allowed by the given electronic level sequence.

The total orbital moments evaluated for the case when the 
magnetic moment is parallel to
the dimer axis (i.e., perpendicular to
the benzene plane) directly reflect the nature
of the HOMO. In the case of Fe$_2$Bz, Fig. \ref{fig:level}, the HOMO is 
a $\pi^*$ state in the majority spin channel. Spin-orbit coupling splits
this state in such a way that the energy of the $m = -1$ sub-level is reduced
($m$ denotes the magnetic quantum number). This sub-level is consequently
occupied, while the $m = +1$ sub-level is empty, and the total orbital
moment is close to $-1 \; \mu_B$.
In the case of Co$_2$Bz, Fig. \ref{fig:level}, the HOMO is
a $\delta^*$ state in the minority spin channel. Spin-orbit coupling splits
this state so that the energy of the $m = +2$ sub-level is reduced.
Accordingly, $\mu_L \approx 2 \; \mu_B$.
In Ni$_2$Bz, the HOMO is a $\delta^*$ state with both sub-levels
occupied. Therefore, the orbital moment nearly vanishes.
Finally, in Ru$_2$Bz the HOMO is a $\delta$ state in the minority spin channel
with $\mu_L \approx 2 \; \mu_B$.

\subsection{\label{sec:mae}MAE} 
After analyzing the spin and the orbital moments in the stable $\perp_{\rm c}$ $\rm
TM_2Bz$ structures, we proceed to another
important property of magnetic systems, the MAE.
This quantity is in the main focus of the present investigation.
In the perpendicular
adsorption mode of $\rm TM_2$ on the benzene molecule, 
it is natural to consider
the MAE as the energy difference between the states with magnetization direction
parallel and perpendicular to the benzene plane, Eq. (\ref{eq:2}).

The first two lines in Table~\ref{tab:mae} list our calculated
MAE for the stable magnetic $\rm TM_2Bz$ structures with TM = Fe, Co, Ni, Ru.
The data for $E_{tot}$
were obtained by two self-consistent fully relativistic calculations with 
respective magnetization directions.
The values calculated
without OP correction should be considered as a lower estimate to the
expected MAE, while an upper estimate is obtained by including the OP correction.   

Endowed with a large ground-state orbital
moment as demonstrated in Sec.~\ref{sec:spin}, $\rm Co_2Bz$ 
and Ru$_2$Bz show a huge MAE.
The lower estimate to the
MAE in $\rm Co_2Bz$ is hardly changed in comparison with the free Co
dimer.\cite{fritsch08}
This is because the
magnetic state and the important features of the electronic structure
of Co$_2$ are not changed by the adsorption. 
The upper estimate, 334 meV per Co$_2$Bz molecule, is even higher
than the related value for Co$_2$ (188 meV per dimer).
This is due to the almost complete localization of
the HOMO on TM2 in the case of Co$_2$Bz (Table~\ref{tab:mae}, lines three and four).
While $\mu_{L({\rm TM2})}^{\perp} \approx 2 \; \mu_B$ in Co$_2$Bz
(Table~\ref{tab:mag}),
it is only half as large in the free Co dimer, where by symmetry
$\mu_{L({\rm TM1})} = \mu_{L({\rm TM2})} \approx 1 \; \mu_B$
in the ground state. The OP
correction energy is only half as large in the latter case compared to the
former one,
since it is quadratic in the atom-projected $\mu_L$.
At variance, the spin-orbit coupling energy is linear in $\mu_L$.

The case of Ru$_2$Bz is different.
We find in the ground state of the free Ru dimer (with $S$ = 2)
a two-fold degenerate, completely occupied majority-spin
$3d-\pi^*$ HOMO and, thus, a small MAE.
The spin moment is reduced by the adsorption of Ru$_2$ on benzene.
This causes a change of the electron configuration, resulting in a
two-fold degenerate and singly occupied $\delta^*$ HOMO with a related huge MAE.

The iron system shows a somewhat smaller MAE, mainly due to the
smaller value of $|m|$ of the HOMO. Finally, the Ni system 
has a fully occupied HOMO that does not contribute to the MAE
in lowest order. Nonetheless, the obtained MAE reaches or exceeds
the highest known experimental values.\cite{gambardella03,milios07}

We finally would like to understand the obtained numbers
in terms of simple arguments based on perturbation theory,\cite{strandberg07}
extended here to include the OP correction.
Our consideration is limited to systems with singly occupied, two-fold
degenerate HOMO (here, Fe$_2$Bz, Co$_2$Bz, and Ru$_2$Bz).
The MAE is approximated by the single-particle energy change of
the occupied HOMO level upon changing the direction of magnetization.
If the OP correction is included, we call the magnetic
anisotropy energy MAE$_{\rm (SO+OP)}$, otherwise it is called MAE$_{\rm (SO)}$.
Thus,
\begin{equation}
{\rm MAE_{(SO)}} \approx \; |m|\; \sum_i\;
(C^2_{m, d-{\rm TMi}} + C^2_{-m, d-{\rm TMi}}) \;
\xi_d/2
\end{equation}
in first-order perturbation theory.
Here, $\xi_d$ is the $d$-shell spin-orbit parameter and 
$C_{m, d-{\rm TM}i}$ is the projection of one of the
HOMO orbitals on the $d$-orbital of atom TM$i$ ($i$ = 1, 2)
with magnetic quantum number $m$ ($|m|$ = 1 or 2 for a HOMO of type $\pi$ or
$\delta$, respectively).
If OP corrections are taken into account, this first-order estimate
changes to
\begin{equation}
{\rm MAE_{(SO+OP)}} \approx
\; |m|\;  \sum_i \; (C^2_{m, d-{\rm TMi}} + C^2_{-m, d-{\rm TMi}})\;
(\xi_d/2 + B \Delta |\mu_{L({\rm TM}i)}|/\mu_B) \;.
\end{equation}
Here, $B$ denotes the TM-specific Racah parameters,\cite{Judd63}
evaluated from the related atomic orbitals, and
$\Delta |\mu_{L({\rm TM}i)}| = |\mu_{L({\rm TM}i)}^{\perp}| -
|\mu_{L({\rm TM}i)}^{\parallel}|$, ($i$ = 1, 2), according to Tab. \ref{tab:mag}.

Table~\ref{tab:mae} shows the major contributions to the
composition of the HOMO, $(C^2_{m, d-{\rm TMi}}+C^2_{-m, d-{\rm TMi}})$,
the related magnetic quantum number $|m|$,
the occupation of the HOMO, the spin-orbit parameters, and the Racah parameters.
MAE values estimated by first-order perturbation theory are given in brackets 
following the self-consistently evaluated MAE data.
We find that the self-consistent values are smaller than the estimates
obtained by perturbation theory. This has at least two reasons:
(i) negative contributions from levels other than the HOMO and higher-order
HOMO contributions, cf. the results
for Ni$_2$Bz where the HOMO does not contribute in first order, and
(ii) charge relaxation reduces the effect.
Nonetheless, the self-consistent MAE amounts to 60 $\ldots$ 80\% of the
first-order estimates.

According to the above analysis,
$\rm Co_2Bz$ and $\rm Ru_2Bz$ are interesting candidates for
strong-MAE applications. It should be noted that in both
cases the easy axis of magnetization is directed perpendicularly to the 
benzene ring. This fact is advantageous for conventional recording techniques.

\section{\label{sec:summary}Summary} 
We report a systematic DFT study of the ground-state structures, bonding
mechanism, spin and orbital moments, and in particular of the MAE of $\rm
TM_2Bz$ complexes (TM=Fe, Co, Ni, Ru, Rh, Pd), using the full-potential
local-orbital method FPLO. Upright adsorption modes with $C_{6v}$ symmetry
of $\rm Fe_2$, $\rm Co_2$, $\rm Ni_2$, and $\rm Ru_2$ on benzene molecules are
confirmed (Fe$_2$Bz) or predicted (others).
Huge MAE, stable geometry, and stable magnetic ground states
are predicted for $\rm Co_2Bz$ and for $\rm Ru_2Bz$.
The main origin of the large anisotropy of
these two systems is the large orbital moment of the TM atom which is farther
away from the benzene plane. Analysis of the electronic states shows that
bonding of the Co dimer on the benzene molecule does not lead to any
deterioration of the magnetic properties of the dimer. Most important
is that the two-fold degenerate singly occupied HOMO state of the free dimer
is preserved, which allows the spin-orbit coupling to produce
a large magnetic anisotropy.
An important conclusion can be drawn from these results: robust and
easy-to-prepare carbon-based substrates are well-suited to adsorb
transition-metal dimers for the purpose of high-density magnetic recording. We
hope that our predicted exceptionally large MAE of $\rm Co_2Bz$ and $\rm
Ru_2Bz$ will motivate experimental investigations of transition-metal dimers
on carbon-based substrates, like graphite or graphene.  

\appendix* 
\section*{\label{sec:appendix}Appendix} 
A full optimization of all atomic positions was carried out for $\rm
Co_2Bz$ complexes by using the ESPRESSO code.\cite{pwscf}
We used the pseudopotentials
Co.pbe-nd-rrkjus.UPF, C.pbe-rrkjus.UPF and H.pbe-van\verb|_|bm.UPF from the
\url{http://www.quantum-espresso.org} distribution. A supercell of the size
20 \AA $\times$ 20 \AA $\times$ 20 \AA ~was used to make sure that
there is virtually no interaction
between the molecules of neighboring cells. The
Brillouin zone sampling was performed only on the $\Gamma$ point. The cutoffs used
for the wave functions and for the charge density were 60 Ry and 300 Ry,
respectively. The convergence in total energy was carefully checked. A 
Marzari-Vanderbilt cold smearing~\cite{marzari99} with 0.007 Ry 
was used to get the convergence in energy levels.
The optimizations were done without any constraints on symmetry
or spin moment. Fig.~\ref{fig:str14} 
shows the 14 initial structures used to start the geometry optimization.   

The optimization results in six types of final structures. Structures (i),
(xi), (xii), and (xiii)
converge to the perpendicular adsorption mode; structure (x) stays almost
unchanged and leads to the parallel adsorption mode on the bridge site
of the carbon ring; structures (v) and
(ix) converge to the parallel adsorption mode on the top site of the carbon
ring; structures (iii), (vi), and (viii) finally go over
to a structure type similar to
(iii); structures (ii) and (iv) become the structure type (iv); and the final
structure of (xiv) is still the adsorption mode on both sides. The optimization
for structure (vii) does not converge. Among these structures, the
perpendicular adsorption mode shows the lowest total energy,
which is 0.74 eV lower than the $\parallel_{\rm b}$ adsorption mode. The spin
of this structure is $S$ = 2, confirming the 
FPLO result. There, the $\parallel_{\rm b}$ adsorption mode was found
0.98 eV higher than the perpendicular adsorption mode.

Full optimization of all atomic positions shows that there is only a tiny
distortion of the benzene plane in the perpendicular adsorption mode. The length
of the C-C bonds increases slightly, two of them changing from 1.40 \AA ~to
1.4175 \AA ~and four of them changing to 1.4169 \AA. 
The C-H bond-length changes from 1.09 \AA ~to 1.0889 \AA ~and 1.0887 \AA.
The electronic structure shows
that the two $\delta^*$ states (now, HOMO and LUMO)
are still nearly degenerate with a gap smaller than 1 meV.

\clearpage

\clearpage

\clearpage

\clearpage
\begin{figure} 
\includegraphics[width=1.0\textwidth]{fig_level_Co_Fe} 
\caption{\label{fig:level}(Color online)
Scalar-relativistic single-particle levels of Co$_2$ (left
panel), Co$_2$Bz (ground-state structure, second panel), benzene (third
panel), Fe$_2$Bz (ground-state structure, fourth panel and first
spin-excited state, fifth panel), 
and Fe$_2$ (right panel). All energies refer to a common vacuum
level. Black lines denote occupied states, orange (gray) lines denote empty states,
and thick blue (gray) lines indicate singly occupied two-fold degenerate states. With
the exception of benzene, the levels are spin-split 
($S$ = 2, 2, 2(3), and 3
for Co$_2$, Co$_2$Bz, Fe$_2$Bz, and Fe$_2$, respectively). Majority
states are indicated by up-arrows, minority states by down-arrows.
Dimer-dominated
states are labeled in black and benzene-dominated states are labeled in red (gray).}
\end{figure}    

\clearpage
\begin{figure} 
\includegraphics[width=1.0\textwidth]{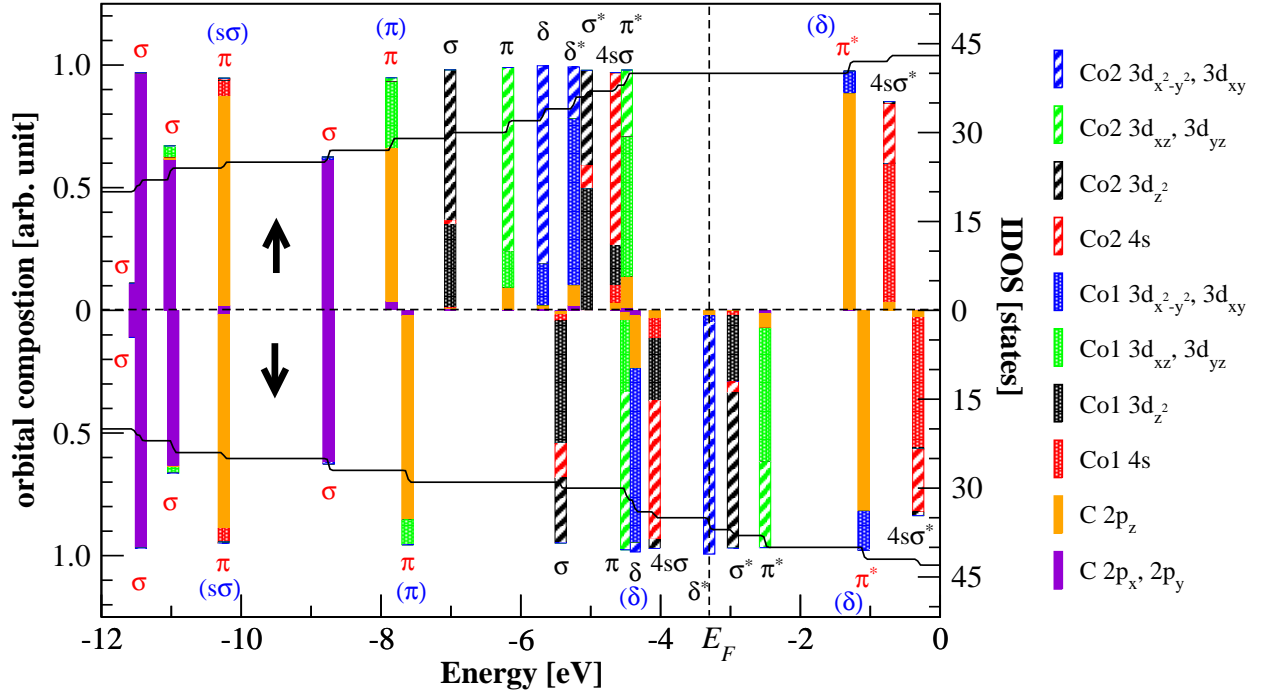} 
\caption{\label{fig:orbcomp}The 
orbital composition of each state (left axis) and the 
integrated density of states (IDOS, right axis)
for the ground-state structure and spin of
$\rm Co_2Bz$. The upward (downward) arrow indicates majority
(minority) spin states.
All energies refer to a common vacuum level. 
The labels for each state are the same as those in Fig.~\ref{fig:level}. 
The three types of chemical bonds between the Co dimer and benzene are
labelled blue in parentheses. 
The Co atom closer to the benzene plane is labelled Co1,
the other one is labelled Co2. The position of the Fermi level ($E_F$) is
indicated by a vertical line. 
Missing contributions to the orbital composition which should add up to
unity are due to the omitted C-$2s$ and H-states.
} 
\end{figure}

\clearpage
\begin{figure} 
\includegraphics[width=1.0\textwidth]{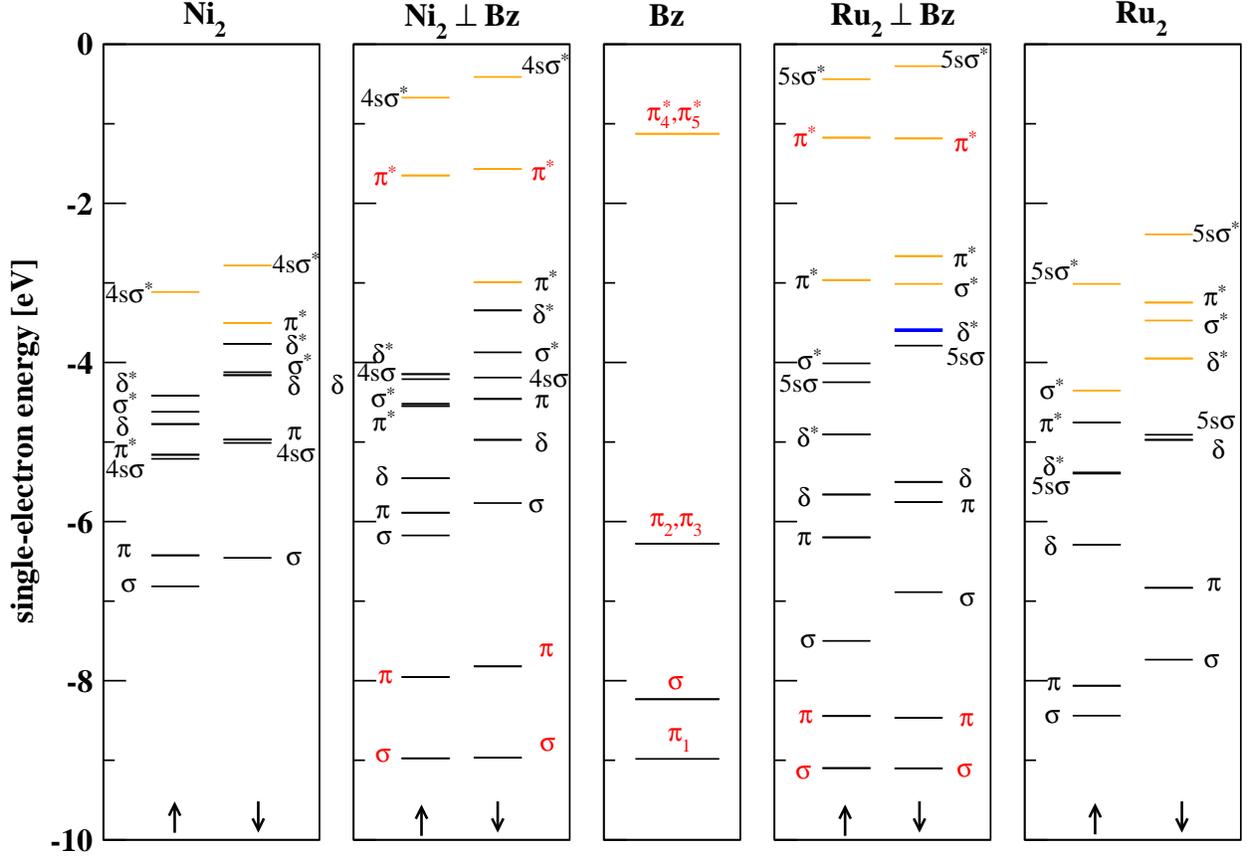} 
\caption{\label{fig:levelniru}(Color online)
Scalar-relativistic single-particle levels of Ni$_2$ (left
panel), Ni$_2$Bz (ground-state structure, second panel), benzene (third
panel), Ru$_2$Bz (ground-state structure, fourth panel),
and Ru$_2$ (right panel). All energies refer to a common vacuum
level. Black lines denote occupied states, orange (gray) lines denote empty states,
and thick blue (gray) lines indicate singly occupied two-fold degenerate states. With
the exception of benzene, the levels are spin-split 
($S$ = 1 in all cases but Ru$_2$, where $S$ = 2).
Majority states are indicated by up-arrows, minority states by down-arrows.
Dimer-dominated
states are labeled in black and benzene-dominated states are labeled in red (gray).}
\end{figure}    

\clearpage
\begin{figure} 
\includegraphics[width=0.84\textwidth]{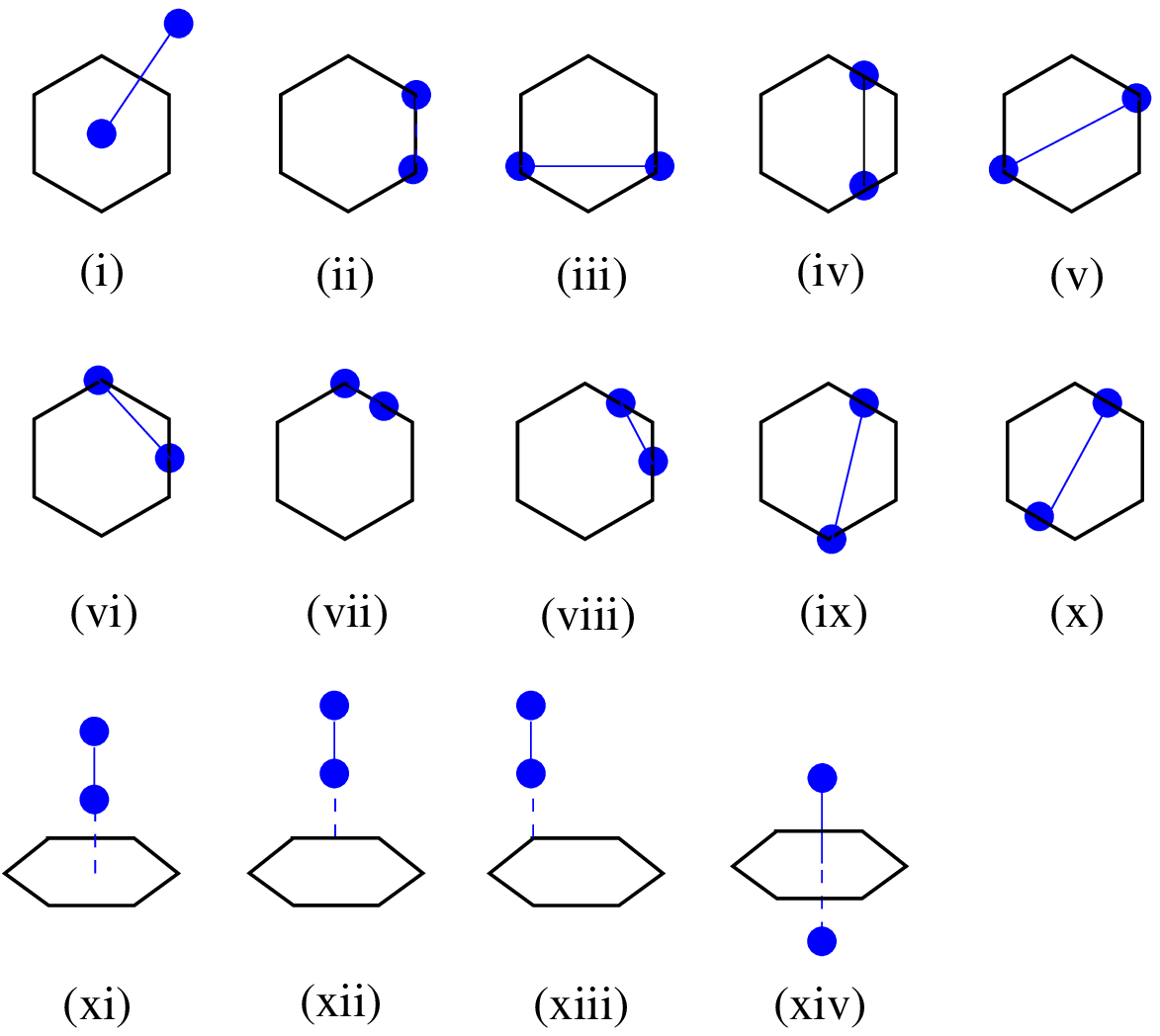} 
\setcounter{figure}{0} 
\renewcommand{\thefigure}{A\arabic{figure}} 
\caption{\label{fig:str14}(Color online) Illustration of 14 initial 
structures optimized by the ESPRESSO code.
From (i) to (x), top-views of possible parallel adsorption modes; from
(xi) to (xiii), side-views of possible upright adsorption modes are shown;
(xiv) is a case in which one Co atom is attached on each side of
the carbon ring.} 
\end{figure}

\clearpage
\begin{table} 
\caption{\label{tab:eb}
Dimer adsorption energies, $E_{ad}$, total and atom-resolved spin magnetic
moments, $\mu_{S\rm (total)}$, $\mu_{S\rm (TM1)}$, and
$\mu_{S\rm (TM2)}$ (TM1 refers to the atom closer to the benzene in case 
of perpendicular bonding, TM2 to the other atom), the distance between the two
transition-metal atoms $d_{\rm TM-TM}$, and the distance between
benzene plane and TM1, $d_{\rm TM-Bz}$,
for the ground-state structures of $\rm TM_2Bz$ (TM = Fe, Co, Ni, Ru, Rh, Pd)
complexes. 
C-C and C-H bond lengths are fixed (1.40 \AA{} and 1.09 \AA{}, respectively).
The structure type of each molecule is labelled according to the notation 
introduced in Fig.~\ref{fig:str}.
Our present results are labelled ``a'',
literature data (for dimers on graphene) are labelled ``b'', Ref.~\onlinecite{johll09}
and ``c'', Ref.~\onlinecite{cabria10}.
}
\begin{ruledtabular}
\begin{tabular}{c @{}d@{}*{10}{d}}
system &\multicolumn{2}{c}{Fe$_2$Bz}&\multicolumn{2}{c}{Co$_2$Bz}&\multicolumn{2}{c}{Ni$_2$Bz}&\multicolumn{1}{c}{Ru$_2$Bz}&\multicolumn{1}{c}{Rh$_2$Bz}&\multicolumn{2}{c}{Pd$_2$Bz}\\ 
\cline{2-3}\cline{4-5}\cline{6-7}\cline{8-8}\cline{9-9}\cline{10-11}
             &\multicolumn{1}{c}{a}&\multicolumn{1}{c}{b}&\multicolumn{1}{c}{a}&\multicolumn{1}{c}{b}&\multicolumn{1}{c}{a}&\multicolumn{1}{c}{b}&\multicolumn{1}{c}{a}&\multicolumn{1}{c}{a}&\multicolumn{1}{c}{a}&\multicolumn{1}{c}{c}\\ 
\hline
structure&\multicolumn{1}{c}{$\perp_{\rm c}$}&\multicolumn{1}{c}{$\perp_{\rm c}$}&\multicolumn{1}{c}{$\perp_{\rm c}$}&\multicolumn{1}{c}{$\perp_{\rm c}$}&\multicolumn{1}{c}{$\perp_{\rm c}$}&\multicolumn{1}{c}{$\perp_{\rm c}$}&\multicolumn{1}{c}{$\perp_{\rm c}$}&\multicolumn{1}{c}{$\parallel_{\rm b}$}&\multicolumn{1}{c}{$\parallel_{\rm b}$}&\multicolumn{1}{c}{$\parallel_{\rm b}$}\\ 
$E_{ad}$(eV)&0.87&0.72&1.39&0.92&1.12&0.96&1.40&0.97&1.16&1.28\\ 
$\mu_{S\rm (total)}$($\mu_B$)&4&6&4&4&2&2&2&2&0&0\\
$\mu_{S\rm (TM1)}$($\mu_B$)&0.75&2.76&1.64&1.66&0.71&0.73&-0.10&1.01&0.00&0.00\\ 
$\mu_{S\rm (TM2)}$($\mu_B$)&3.35&3.48&2.45&2.43&1.29&1.29&2.13&1.01&0.00&0.00\\ 
$d_{\rm TM-TM}$(\AA)&2.04&2.08&2.09&2.03&2.15&2.14&2.22&2.49&2.80&2.75\\ 
$d_{\rm TM-Bz}$(\AA)&1.60&1.86&1.66&1.72&1.71&1.73&1.77&2.08&2.14&2.15\\ 
\end{tabular}
\end{ruledtabular}
\end{table}  

\begin{table} 
\caption{\label{tab:mag}Spin moments $\mu_S$ and orbital moments
$\mu_L$ (in $\mu_B$) for the ground-state structures of $\rm
TM_2Bz$ (TM = Fe, Co, Ni, Ru) complexes calculated within
the fully relativistic scheme
with magnetization perpendicular ($\perp$) or parallel ($\parallel$) to the
benzene plane. The effect of the OP correction is also illustrated
by comparing the
values calculated without the OP correction (SO) and with the OP correction
(SO+OP).
} 
\begin{ruledtabular}
\begin{tabular}{c@{} @{}d@{}*{8}{d}} 
&\multicolumn{2}{c}{$\rm Fe_2Bz$}&\multicolumn{2}{c}{$\rm Co_2Bz$}&\multicolumn{2}{c}{$\rm
Ni_2Bz$}&\multicolumn{2}{c}{$\rm Ru_2Bz$}\\
\cline{2-3}\cline{4-5}\cline{6-7}\cline{8-9} 
&\multicolumn{1}{c}{SO}&\multicolumn{1}{c}{SO+OP}&\multicolumn{1}{c}{SO}&\multicolumn{1}{c}{SO+OP}&\multicolumn{1}{c}{SO}&\multicolumn{1}{c}{SO+OP}&\multicolumn{1}{c}{SO}&\multicolumn{1}{c}{SO+OP}\\ 
\hline
$\mu_{S\rm (TM1)}^{\perp}$&0.75&0.73&1.64&1.64&0.71&0.71&-0.09&-0.10\\ 
$\mu_{S\rm (TM1)}^{\parallel}$&0.75&0.75&1.64&1.64&0.71&0.67&-0.09&-0.09\\ 
$\mu_{L\rm (TM1)}^{\perp}$&-0.72&-0.89&0.07&-0.10&0.01&0.02&0.04&-0.04\\ 
$\mu_{L\rm (TM1)}^{\parallel}$&0.02&0.05&0.13&0.39&0.04&-0.31&0.03&0.02\\ 
\hline
$\mu_{S\rm (TM2)}^{\perp}$&3.34&3.36&2.45&2.46&1.28&1.28&2.13&2.13\\ 
$\mu_{S\rm (TM2)}^{\parallel}$&3.34&3.34&2.45&2.45&1.28&1.33&2.10&2.11\\ 
$\mu_{L\rm (TM2)}^{\perp}$&-0.17&-0.17&1.93&2.12&0.01&0.00&1.91&2.01\\ 
$\mu_{L\rm (TM2)}^{\parallel}$&0.11&0.25&0.17&0.54&0.37&1.69&0.11&0.21\\
\hline
$\mu_{L\rm (total)}^{\perp}$&-0.89&-1.06&2.00&2.02&0.02&0.02&1.95&1.97\\
$\mu_{L\rm (total)}^{\parallel}$&0.13&0.30&0.30&0.93&0.41&1.38&0.14&0.23\\ 
\end{tabular}
\end{ruledtabular}
\end{table}  

\begin{table} 
\caption{\label{tab:mae}The MAE (per molecule),
calculated using Eq. (\ref{eq:2}) for the
ground-state structures of $\rm TM_2Bz$ (TM = Fe, Co, Ni, Ru). Positive values
of MAE indicate that the easy axis of the system is perpendicular to the
benzene plane, while negative values mean that the direction parallel to the
benzene plane is the easy axis. Both a lower estimate of the MAE calculated without OP
correction ($\rm MAE_{(SO)}$) and an upper estimate of the MAE obtained with OP
correction ($\rm MAE_{(SO+OP)}$) are listed. Data in brackets indicate 
estimates obtained by first-order perturbation theory, see text.
Further, the principal composition $C^2_{m, d-{\rm TMi}} + C^2_{-m, d-{\rm TMi}}$,
the magnetic quantum number $|m|$,
and the occupation of the HOMO are given as well as the
spin-orbit coupling parameter $\xi_d$ and the Racah parameter $B$.
} 
\begin{ruledtabular}
\begin{tabular}{l @{}*{4}{c}} 
&\multicolumn{1}{c}{$\rm Fe_2Bz$}&\multicolumn{1}{c}{$\rm Co_2Bz$}&\multicolumn{1}{c}{$\rm Ni_2Bz$}&\multicolumn{1}{c}{$\rm Ru_2Bz$}\\
\hline
$\rm MAE_{(SO)}$ (meV)&+15 [+25] &+51 [+74] &$-7$&+104 [+123] \\ 
$\rm MAE_{(SO+OP)}$ (meV)&+61 [+107] &+334 [+519] &$-96$&+279 [+403] \\ 
$C^2_{m, d-{\rm TM1}} + C^2_{-m, d-{\rm TM1}}$ & 68\% & 3\% & 21\% & 2\% \\
$C^2_{m, d-{\rm TM2}} + C^2_{-m, d-{\rm TM2}}$ & 15\% & 94\% & 77\% & 94\% \\
$|m|$ of the HOMO & 1 & 2 & 2 & 2 \\
occupation of the HOMO&1&1&2&1\\
$\xi_d$ (meV)&61&76&96&128\\
$B$ (meV) & 140 & 149 & 154 & 95 \\
\end{tabular}
\end{ruledtabular}
\end{table}

\clearpage
\begin{acknowledgments}
Discussions with Helmut Eschrig and with
Hway Chuan Kang are gratefully acknowledged.
\end{acknowledgments}

\clearpage

%

\end{document}